\documentclass[english,12pt,a4paper]{article}
\pdfoutput=1
\usepackage{authblk}
\usepackage[text={17.0cm,23.7cm}]{geometry}
\usepackage[T1]{fontenc}
\usepackage[utf8]{inputenc}
\usepackage[english]{babel} 
\usepackage{hyperref}
\usepackage{graphicx}
\usepackage{amsmath}
\usepackage{amssymb}
\usepackage[dvipsnames]{xcolor}
\usepackage{yfonts}
\usepackage{ulem}
\usepackage{enumitem}
\usepackage[mathscr]{euscript}
\usepackage[style=numeric-comp, sorting=none]{biblatex}
\usepackage{float}
\usepackage{amsmath}
\usepackage{csquotes}

\begin{document}
	
	\title{\vspace{-2cm}
		{\normalsize
			\flushright TUM-HEP 1418/22\\}
		\vspace{0.6cm}
	\textbf{New constraints on the dark matter-neutrino and dark matter-photon scattering cross sections from TXS 0506+056}\\[8mm]}
\author[1]{Francesc Ferrer}
\author[2,3]{Gonzalo Herrera}
\author[2]{Alejandro Ibarra}
\affil[1]{\normalsize\textit{{Department of Physics and McDonnell Center for the Space Sciences, Washington University, St. Louis, MO 63130, USA}}}
\affil[2]{\normalsize\textit{Physik-Department, Technische Universit\"at M\"unchen, James-Franck-Stra\ss{}e, 85748 Garching, Germany}}
\affil[3]{\normalsize\textit{Max-Planck-Institut f\"ur Physik (Werner-Heisenberg-Institut), F\"ohringer Ring 6,80805 M\"unchen, Germany}}

\date{}
\maketitle

\maketitle
\begin{abstract}
	The flux of high energy neutrinos and photons produced in a blazar could get attenuated when they propagate through the dark matter spike around the central black hole and the halo of the host galaxy. Using the observation by IceCube of a few high-energy neutrino events from TXS 0506+056, and their coincident gamma ray events, we obtain new constraints on the dark matter-neutrino and dark matter-photon scattering cross sections. Our constraints are orders of magnitude more stringent than those derived from considering the attenuation through the intergalactic medium and the Milky Way dark matter halo. When the cross-section increases with energy, our constraints are also stronger than those derived from the CMB and large-scale structure.
\end{abstract}

\section{Introduction}
\label{sec:Intro}

High-energy particles are produced in astrophysical sources and can reach the Earth, providing valuable information about the environment where these particles have been produced and the medium through which they have propagated. While high-energy photons have been detected and their sources had been identified long ago, it is only very recently that sources of high energy neutrinos have been discovered. On 22 September 2017, the IceCube neutrino observatory detected a neutrino event with an energy of 290 TeV, consistent with the direction of the gamma-ray blazar TXS 0506+056, located at a distance of 1421 Mpc~\cite{IceCube95,2017ATel10791....1T,}. The neutrino alert from IceCube triggered an observation campaign ranging from radio to gamma-ray telescopes~\cite{IceCube:2018dnn}. In particular, Fermi-LAT observed an excess of gamma rays from the direction of TXS 0506+056 following the IceCube alert, with more than 5$\sigma$ significance and reaching energies up to 300 GeV. On the other hand, the initial observations of TXS 0506+056 by ground based gamma-ray telescopes after the IceCube alert only lead to an upper bound on the gamma-ray flux. Nevertheless, a few days later MAGIC detected high-energy gamma-rays up to 400 GeV with a significance that reached  5$\sigma$ after a few hours of observation, and these observations are compatible with the upper limits from HESS and VERITAS. Several works have shown that the observed neutrino and gamma-ray fluxes from TXS 0506+056 can be explained with leptohadronic models, where the high energy neutrinos are produced mainly via $pp$ and $p\gamma$ processes, and the bulk of gamma-rays is produced via leptonic processes, with a non-negligible contribution from hadronic processes \cite{Cerruti_2018, Keivani:2018rnh,Murase:2018iyl,Gao:2018mnu, Xue_2019,Zhang:2019htg, Petropoulou_2020, Capel:2022cnm}.

The gamma ray flux from distant sources is subject to attenuation due to electromagnetic processes in the blazar jet, as well as during their propagation to the Earth due to its interactions with the extragalactic background light, {\it e.g.} via the process $\gamma \gamma \rightarrow e^{+}e^{-}$. Neutrinos on the other hand interact very weakly with matter, and therefore they are generically expected to suffer less attenuation during their propagation. Nonetheless, in extensions of the Standard Model, there could be new gamma-ray or neutrino interactions which could affect their propagation to the Earth~\cite{Mikheyev:1985zog, Athar:2000yw,Mena:2014sja, Bustamante_2015, Denton:2018aml,Murase:2019xqi}. A notable example arises in scenarios where the dark matter of the Universe is constituted by new elementary particles which interact with the photon or the neutrino. 
Therefore, the observation of neutrinos from TXS 0506+056 by IceCube can be used to set constraints on the dark matter-neutrino scattering cross-section, from the requirement that the neutrino flux is not significantly attenuated due to interactions with the dark matter during their propagation to the Earth (for previous works considering attenuation in the intergalactic medium and the Milky Way, see \cite{Choi:2019ixb, Kelly:2018tyg, Alvey_2019,Arg_elles_2017}). Likewise, the observation of photons by MAGIC and Fermi-LAT can be used to set constraints on the dark matter-photon scattering cross-section.

In this work we consider the attenuation of the gamma-ray and the neutrino flux {\it within} the host galaxy of TXS-0506+056. Importantly, the supermassive black hole at the center of the blazar is expected to seed the formation of a dark matter spike that extends from $\sim 10^{-4}$ pc to $\sim 1$ pc, where the density of dark matter particles is substantially larger than the one expected from a naive extrapolation of the galactic density profile. Moreover, it was estimated in \cite{Padovani_2019} that the bulk of the high-energy neutrinos and gamma-rays from TXS 0506+056 are emitted from a region that is close to the Broad Line Region of the blazar $R_{\rm BLR} \sim 0.021$ pc, which lies within the TXS 0506+056 dark matter spike. Therefore, these particles must traverse the spike and the dark matter halo of the host galaxy (and possibly get scattered) before leaving to the intergalactic medium. In this paper we will argue that  the absorption of the gamma-ray or neutrino fluxes in the spike can be significant, despite its small size, due to the high density of dark matter particles, and we will derive new limits on the scattering cross-section of dark matter particles with photons or neutrinos.

The paper is organized as follows. In Section \ref{sec:DMspike}, we will present the dark matter density profile around TXS 0506+056, and we will calculate the column density encountered by a high energy neutrino or photon emitted by the blazar on its way out of the host galaxy.  In Section \ref{sec:Constraints}, we will set upper limits on the dark matter-neutrino and dark matter-photon cross sections from requiring that the fluxes are not significantly attenuated. Finally, in Section \ref{sec:conclusions}, we will present our conclusions.

\section{Flux attenuation in the vicinity of TXS 0506+056}\label{sec:DMspike}

The dark matter in the vicinity of a black hole generically forms a dense spike \cite{Quinlan:1994ed,Gondolo_1999}. Assuming that the growth of the black hole is adiabatic, an initially cuspy dark matter profile of the form $\rho (r) = \rho_0 (r/r_0)^{-\gamma}$ evolves into:
\begin{align}
	\rho_{\rm sp}(r) = \rho_{R} \, g_{\gamma}(r)\, \Big(\frac{R_{sp}}{r}\Big)^{\gamma_{\rm sp}}\;,
\end{align}
where $R_{\rm sp}=\alpha_{\gamma}r_0(M_{\rm BH}/(\rho_{0}r_{0}^{3})^{\frac{1}{3-\gamma}}$ is the size of the spike, with and $\alpha_\gamma$  a complicated function of $\gamma$ (numerically $\alpha_\gamma\approx 0.1$ for $\gamma=0.7-1.4$, \cite{Gondolo_1999}), and $\gamma_{\rm sp}=\frac{9-2\gamma}{4-\gamma}$ parametrizes the cuspiness of the spike. Further, $g_{\gamma}(r)$ is a function which  can be approximated for $0<\gamma <2 $ by  $g_{\gamma}(r) \simeq (1-\frac{4R_{S}}{r})$, with $R_S$ the Schwarzschild radius, while $\rho_{\rm R}$ is a normalization factor, chosen to match the density profile outside of the spike, $\rho_R=\rho_{0}\, (R_{sp}/r_0)^{-\gamma}$. This density profile is defined only for $r\gtrsim 4 R_S$; for smaller radial coordinates, the density profile vanishes.

In the following, we will assume that far away from the black hole, the dark matter distribution follows the standard NFW profile~\cite{Navarro:1996gj,Navarro:1995iw}, which scales as $\gamma$=1 in the central region, resulting in a spike with $\gamma_{\rm sp}=7/3$ and $\alpha_{\gamma}\simeq 0.1$. The mass of the supermassive black hole at the center of the blazar TXS 0506+056 was estimated in \cite{Padovani_2019} to be $M_{\rm BH}\approx 3 \times 10^{8} M_{\odot}$, so that $R_S\approx 3.0 \times 10^{-5}$ pc. We have taken $r_0$=10 kpc, typical of galaxies hosting BL Lac objects, with a similar size as the Milky Way, for example, for which $r_0\sim 20$ kpc. Finally, the normalization $\rho_0$ is determined by the uncertainty on the black hole mass \cite{Gorchtein_2010,Lacroix_2017}. We find the value $\rho_0\simeq7\times 10^3$ GeV/cm$^{3}$. 

Strictly speaking, this profile only holds when dark matter particles  do not annihilate ({\it e.g} as in scenarios of asymmetric dark matter), or do so very slowly. Otherwise, the maximal dark matter density in the inner regions of the spike is saturated to $\rho_{\text {sat }} = m_{\rm DM} /(\langle\sigma v \rangle t_{\mathrm{BH}})$, where $\langle \sigma v \rangle$ is the velocity averaged dark matter annihilation cross section, and $t_{\rm BH}$ is the time elapsed since the black hole formation, for which we take the value $t_{\rm BH}=10^9$ yr \cite{Wang:2021jic}. Further, the dark matter profile of the spike extends to a certain maximal radius $R_{\rm sp}$, beyond which the dark matter distribution follows the pre-existing NFW profile. In full generality, the dark matter profile in the spike reads \cite{Gondolo_1999} (see also \cite{Lacroix_2015, Lacroix_2017})
\begin{align}\rho(r)= \begin{cases} 
		0 & r\leq 4R_S \\
		\frac{\rho_{\rm sp}(r)\rho_{\rm sat}}{\rho_{\rm sp}(r)+\rho_{\rm sat}} & 4R_S\leq r\leq R_{sp} \\
		\rho_{0}\Big(\frac{r}{r_0}\Big)^{-\gamma} \Big(1+\frac{r}{r_0}\Big)^{-3+\gamma} & r\geq R_{sp} .
	\end{cases}
	\label{eq:spike_profile}
\end{align}

\begin{figure}[t]
		\centering
		\includegraphics[width=0.49\textwidth]{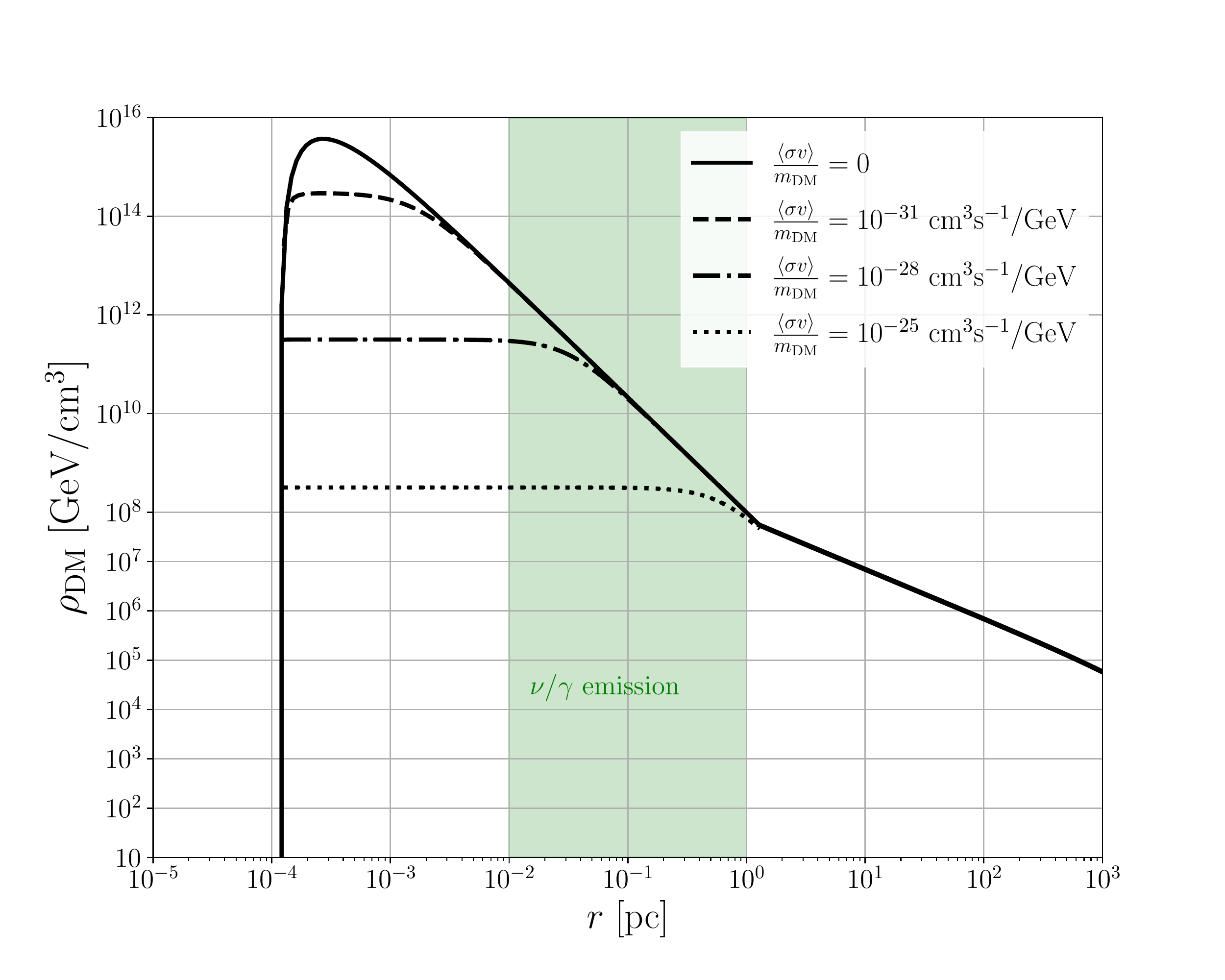}
		\includegraphics[width=0.49\textwidth]{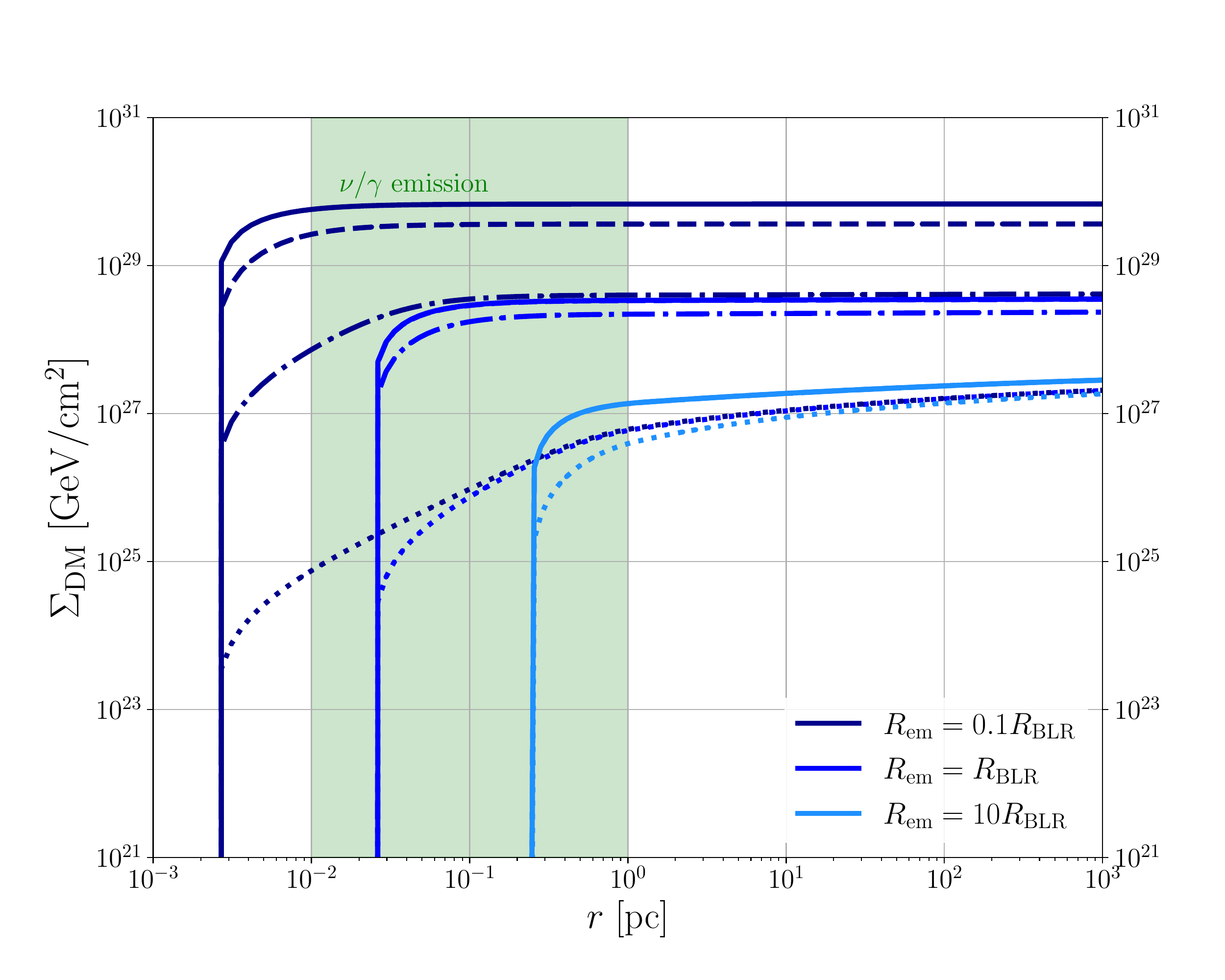}
		\centering
		\caption{{\it Left panel}: dark matter distribution around the black hole of TXS 0506+056, for different values of the dark matter annihilation cross section over its mass. The green shaded region indicates the range of values considered in this work for the emission region of high-energy neutrinos and gamma rays. {\it Right panel}: Total dark matter mass along the line of sight to the emission region of high-energy neutrinos and gamma-rays in TXS 0506+056, in terms of the radius of the broad line region of the blazar ($R_{\rm BLR}\simeq 0.023$ pc), for the halo profiles shown in the left panel.}
		\label{fig:DMspike}
	\end{figure}

The dark matter profile of TXS 0506+056 is shown in the left panel of  Figure \ref{fig:DMspike} for various values of $\langle \sigma v\rangle/m_{\rm DM}$. As apparent from the plot, the dark matter density is extremely high at the position of the broad line region $R_{\rm BLR}\sim 0.023$ pc, where neutrinos and photons are likely to be produced  \cite{Padovani_2019}, and interactions with dark matter particles may occur with sufficient frequency to produce a sizable attenuation of the flux. In order to be conservative, in our work we will also allow for neutrino/photon emission at larger distances from the black hole, where the density is lower. Concretely, we will consider the range 
$R_{\rm em}=10^{-2}-1$ pc for the region of the blazar jet where neutrinos and gamma-rays are produced, indicated in the Figure as a green region.

Let us note that a more accurate treatment of the adiabatic growth of the dark matter spike including relativistic effects, shows that in fact the spike vanishes at $r=2 R_{S}$ instead of $4 R_{S}$, and that the density of dark matter particles is significantly boosted near the core \cite{Sadeghian_2013}. This enhancement is even more pronounced for a rotating black hole  \cite{Ferrer:2017xwm}. On the other hand, the difference with respect to Eq.~(\ref{eq:spike_profile}) is only significant close to the Schwarzschild radius, at $r \lesssim 10^{-3}$ pc, whereas the photons and neutrinos are produced further out.
We will then disregard these relativistic effects in our analysis, and we will use the profile Eq.~(\ref{eq:spike_profile}).

The flux of neutrinos and photons produced at the distance $R_{\rm em}$ from the black hole gets attenuated due to interactions with the medium on their way to the Earth
\begin{align}
\frac{\Phi^{\rm obs}_i}{\Phi^{\rm em}_i}= e^{-\mu_i}
\end{align}
where $\Phi^{\rm obs}_i$ and $\Phi^{\rm em}_i$ are respectively the observed and emitted fluxes of the particle $i$ ($i=\nu$ or $\gamma$),
and $\mu_i$ is an attenuation coefficient that receives contributions from scatterings with Standard Model particles (photons, protons, etc.) as well as from dark matter particles. The attenuation due to dark matter reads:
\begin{align}
\mu_{i}\big|_{\rm DM}=\frac{\sigma_{\rm DM-i}}{m_{\rm DM}} \Sigma_{\rm DM}
\end{align}
where $\sigma_{\rm DM-i}$ is the scattering cross section of dark matter with the particle $i$, and $\Sigma_{\rm DM}$ is the column density of dark matter particles along the path of the particle $i$:
\begin{align}
\Sigma_{\rm DM}=\int_{\rm path} dr\rho(r)
\label{eq:Sigma}
\end{align}

In this paper we focus on the impact on the attenuation of the passage through the dark matter in TXS 0506+056, with density profile given in Eq.~(\ref{eq:spike_profile}), and that as we will see later it is orders of magnitude stronger than the contribution to $\Sigma_{\rm DM}$ from the dark matter in the intergalactic medium and in the Milky Way.
We will then approximate:
\begin{align}
\Sigma_{\rm DM}\simeq 
\Sigma_{\rm DM}\Big|_{\rm spike} +
\Sigma_{\rm DM}\Big|_{\rm host}\simeq \int_{R_{\rm em}}^{R_{\rm sp}} dr\rho(r)+\int_{R_{\rm sp}}^\infty dr\rho(r)\;.
\end{align}

To calculate $\Sigma_{\rm DM}|_{\rm spike}$ we note that in the region where neutrinos and gamma rays are produced $ R_{\rm em}\gg 4 R_{\rm S}$, therefore $g_\gamma(r) \simeq 1$.  When the annihilation cross-section is very small, the dark matter density in the emission region is much smaller than the saturation density (see the left panel of Fig.~\ref{fig:DMspike}). Then, the density profile in this region reads $\rho_{\rm sp}(r)\simeq\rho_{\rm sp}(R_{\rm em})(\frac{r}{R_{\rm em}})^{-\gamma_{\rm sp}}$ and we can write:
\begin{align}
	\Sigma_{\rm DM}\big|_{\rm spike}& \simeq \int_{R_{\rm em}}^{R_{\rm sp}} dr \rho_{\rm sp}(R_{\rm em})\left(\frac{r}{R_{\rm em}}\right)^{-\gamma_{\rm sp}}
	\simeq \frac{\rho_{\rm sp}(R_{\rm em}) R_{\rm em}}{(\gamma_{\rm sp}-1)} \left[1-\left(\frac{R_{\rm sp}}{R_{\rm em}}\right)^{1-\gamma_{\rm sp}}\right]\;.
	\label{eq:SigmaSpike}
\end{align}
As expected, in this regime $\Sigma_{\rm DM}|_{\rm spike}$ is fairly insensitive to the annihilation cross-section, since annihilations occur at a small rate within the emission region, and the profile in this region is practically indistinguishable from the case with $\langle \sigma v \rangle=0$.
On the other hand, when the cross-section is very large, the dark matter density is approximately equal to the saturation density. Then, 
\begin{align}
	\Sigma_{\rm DM}\big|_{\rm spike}& \simeq \int_{R_{\rm em}}^{R_{\rm sp}} dr \rho_{\rm sat}
	\simeq \rho_{\rm sat} R_{\rm sp} \left[1-\frac{R_{\rm em}}{R_{\rm sp}}\right]\;,
	\label{eq:Sigma_large_x-section}
\end{align}
which is inversely proportional to $\langle \sigma v \rangle/m_{\rm DM}$. In general, 
\begin{align}
\Sigma_{\rm DM}\big|_{\rm spike}&=\int_{R_{\rm em}}^{R_{\rm sp}} dr \frac{\rho_{\rm sp}(r)\rho_{\rm sat}}{\rho_{\rm sp}(r)+\rho_{\rm sat}} 
\simeq \frac{\rho_{\rm sp}(R_{\rm em}) R_{\rm em}}{(\gamma_{\rm sp}-1)}
\Big[f(1)-f\left(\frac{R_{\rm sp}}{R_{\rm em}}\right)\Big]\;,
\label{eq:Sigma_general}
\end{align}
where 
\begin{align}
f(x)= x^{1-\gamma_{\rm sp}} {\,}_{2} F_{1}\left(1,1-\frac{1}{\gamma_{\rm sp}},2-\frac{1}{\gamma_{\rm sp}};-\frac{\rho_{\rm sp}(R_{\rm em})}{\rho_{\rm sat}} x^{-\gamma_{\rm sp}}\right) \;,
\label{eq:f}
\end{align}
with ${\,}_{2} F_{1}(a,b;c;z)$ the hypergeometric function .

Further, the contribution to $\Sigma_{\rm DM}$ from the passage through the halo of the host galaxy is:
\begin{align}
	\Sigma_{\rm DM}\Big|_{\rm host}=
	\int_{R_{\rm sp}}^{\infty}dr\;\rho_{0}\Big(\frac{r}{r_0}\Big)^{-1} \Big(1+\frac{r}{r_0}\Big)^{-2}\simeq \rho_0 r_0 \Big[\log\left(\frac{r_0}{R_{\rm sp}}\right)-1\Big]\;,
\end{align}
where we have used $r_0\gg R_{\rm sp}$. This contribution cannot be neglected. First, the dark matter density is still very large in the proximity of the spike (this is in contrast to the path of the neutrinos or photons from TXS 0506+056 through the Milky Way halo on its way to the Earth, which never gets that close to the Galactic center). Second, the dark matter halo extends for several tens of kpc, which can compensate for the smaller dark matter density. 

We show in the right panel of Fig. \ref{fig:DMspike} the value of $\Sigma_{\rm DM}$ as a function of the distance from the black hole for three different values of the distance of the emitting region of neutrinos or gamma-rays ($R_{\rm em}=0.1, 1, 10\,R_{\rm BLR}$) and for the halo profiles considered in the left panel of the figure, sampling different values of the dark matter annihilation cross-section over its mass.

	\section{Constraints on the dark matter-neutrino and dark matter-photon cross section}\label{sec:Constraints}
	
	To derive upper bounds on the interaction cross section of neutrinos with dark matter, we follow the same procedure as in~\cite{Choi:2019ixb, Kelly:2018tyg, Alvey_2019}. Namely, we impose that the attenuation due to dark matter-neutrino interactions is less than 90\%, which translates into $\mu_\nu|_{\rm DM} \lesssim 2.3$. The attenuation of the photon flux is more uncertain, since photons interact more strongly than neutrinos with Standard Model particles in the medium.
	Therefore, we will impose a more aggressive criterion and we will require that the attenuation of the photon flux due to dark matter interactions is less than 99\%, {\it i.e.} $\mu_\gamma|_{\rm DM}\lesssim 4.6$.
	These requirements on the attenuation then allow to set upper limits on the scattering cross section over the mass:
	\begin{align}
		\frac{\sigma_{\rm DM-\nu}}{m_{\mathrm{DM}}} & \lesssim \frac{2.3}{\Sigma_{\rm DM}}\;, \nonumber \\
		\frac{\sigma_{\rm DM-\gamma}}{m_{\mathrm{DM}}} & \lesssim \frac{4.6}{\Sigma_{\rm DM}}\;,
		\label{eq:criteria}
	\end{align}
	with $\Sigma_{\rm DM}$ given in Eq.~(\ref{eq:Sigma}) (see also Fig.~\ref{fig:DMspike}).

\begin{figure}[t!]
		\centering
		\includegraphics[width=0.475\textwidth]{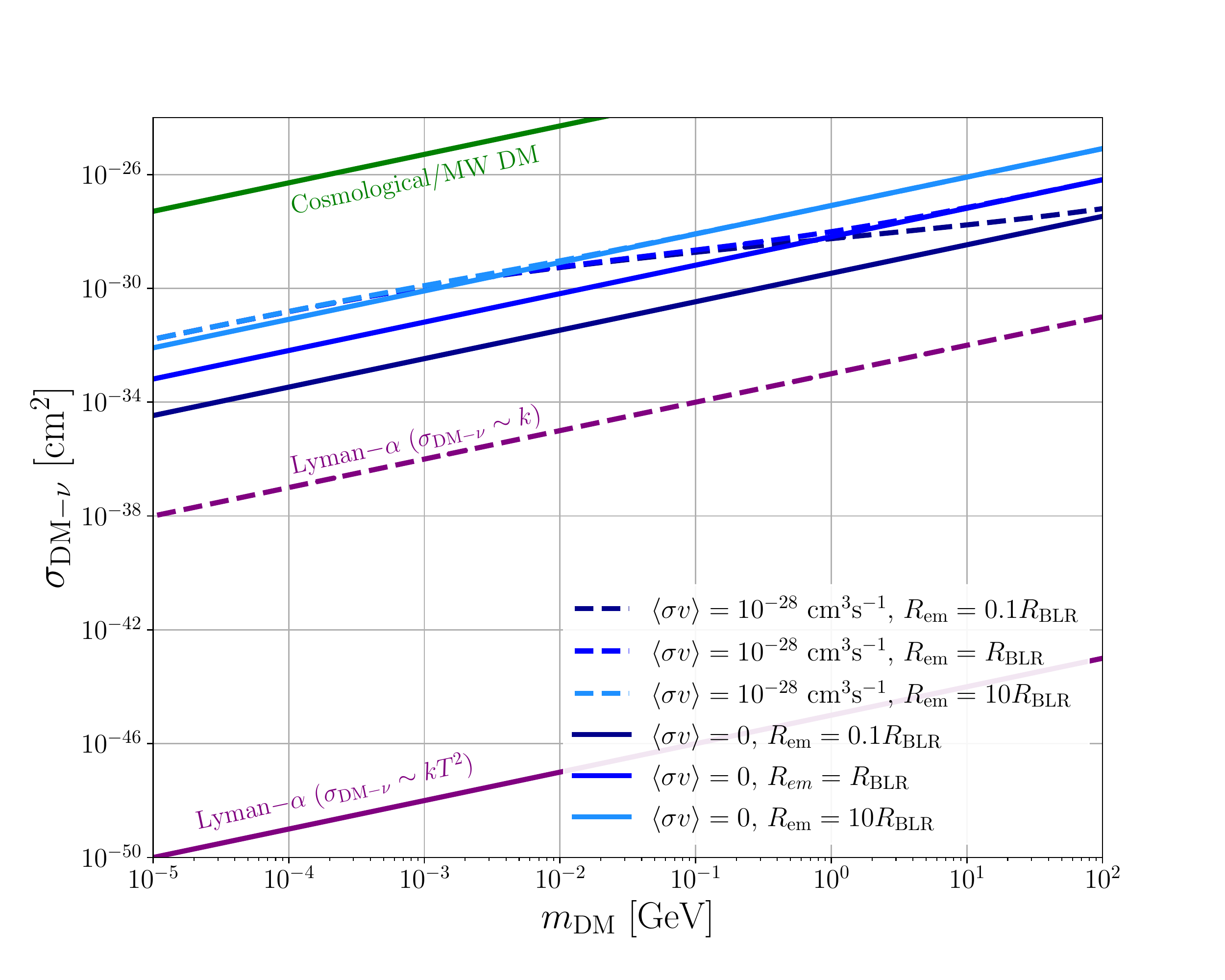}
		\includegraphics[width=0.475\textwidth]{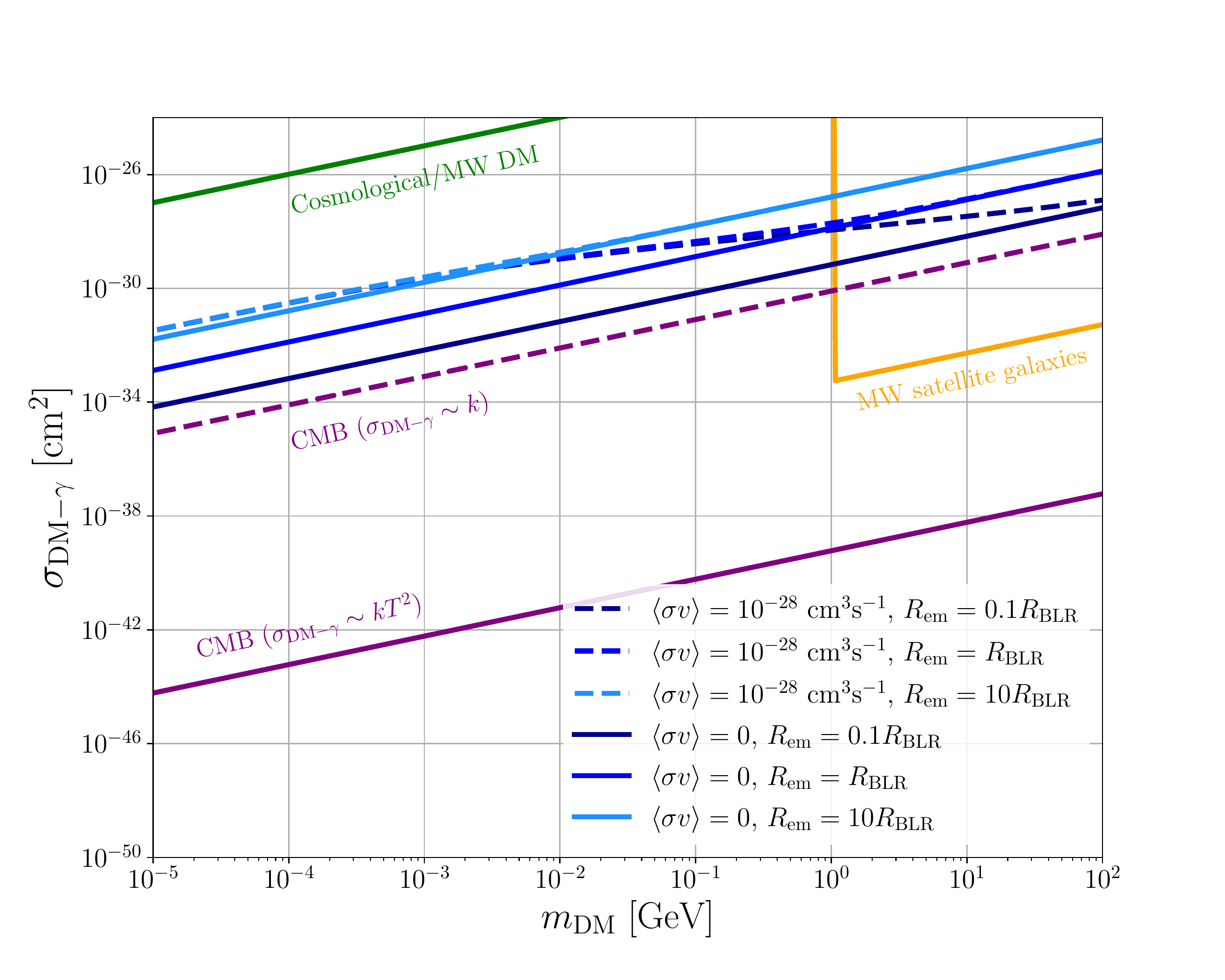}	
	\includegraphics[width=0.475\textwidth]{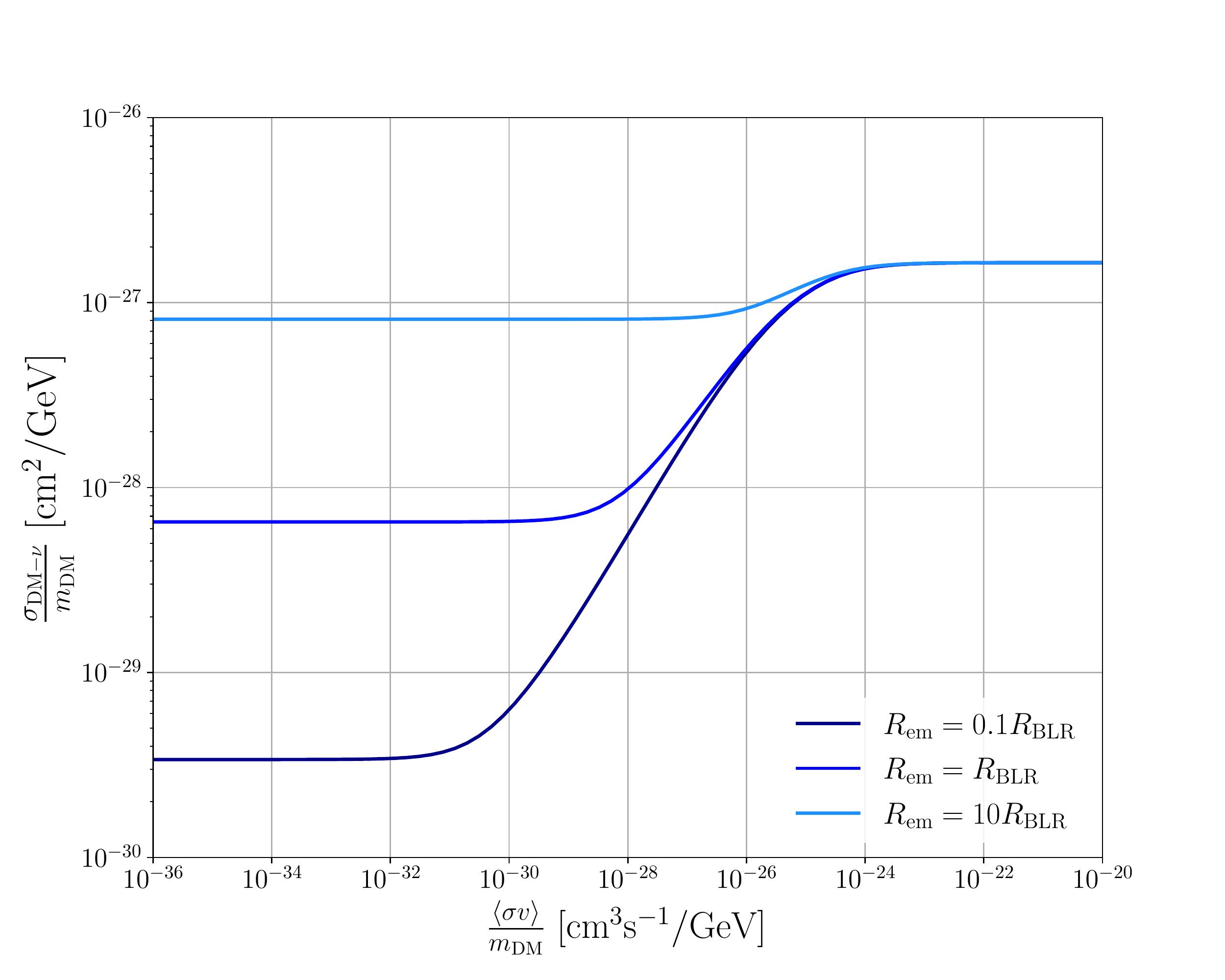}
	\includegraphics[width=0.475\textwidth]{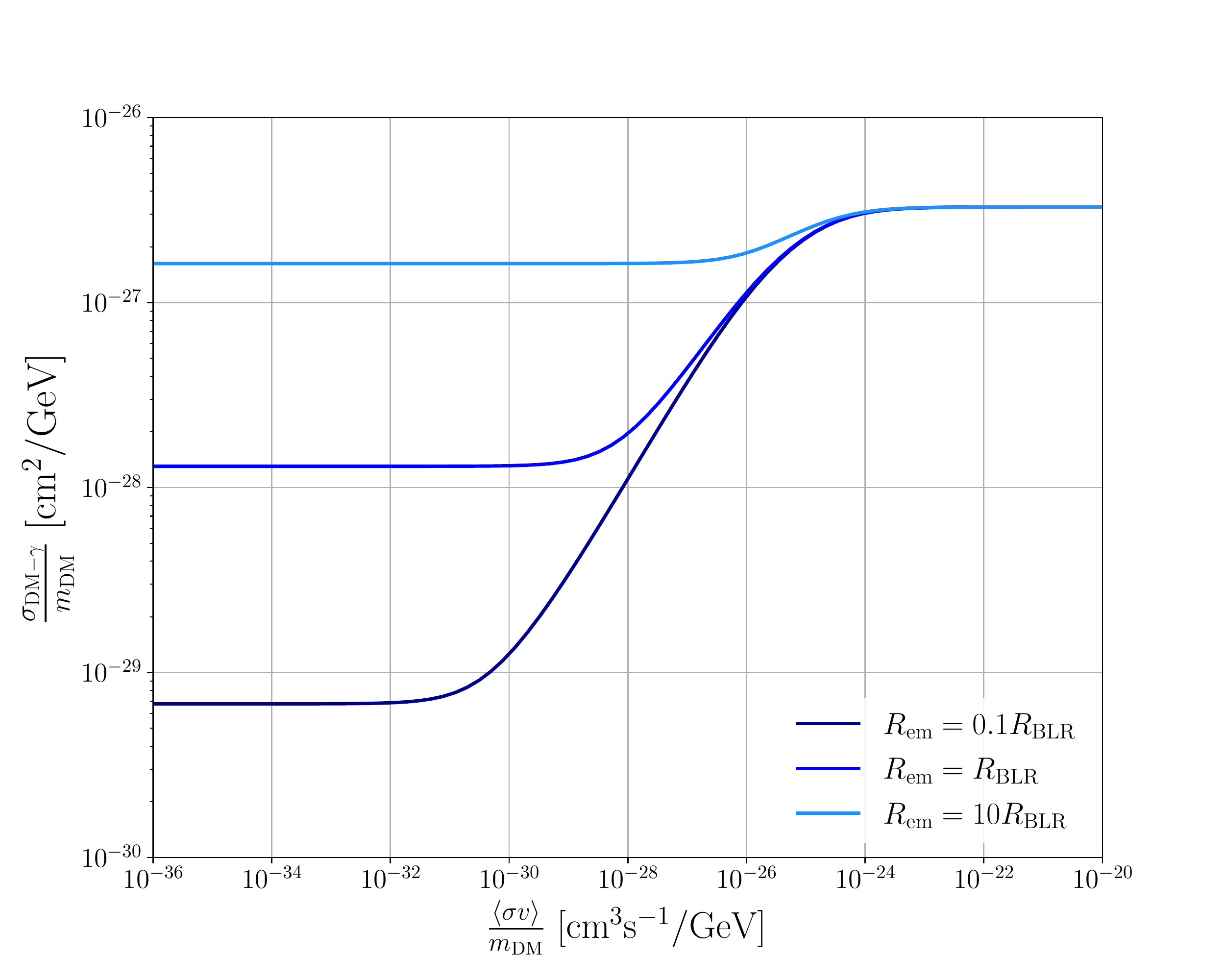}
		\centering
		\caption{{\it Top panels:} Upper limits on the dark matter-neutrino (left panel) and dark matter-photon (right panel) scattering cross-sections as a function of the dark matter mass, derived from the requirement that the neutrino (photon) flux is attenuated by less than 90\% (99\%) when traversing the dark matter spike and the
		galactic halo of TXS 0506+056. The different blue shadings correspond to different locations of the neutrino or photon emission, while the different dashing denote different spike profiles. The green line shows the limits derived from the passage of the neutrinos or photons through the intergalactic medium and the Milky Way halo, the purple lines from Lyman-$\alpha$ observations, and the orange line from Milky Way satellite galaxy counts. {\it Bottom panels:} Dependence on the limits on $\sigma_{\rm DM-\nu}/m_{\rm DM}$ (left panel) and $\sigma_{\rm DM-\gamma}/m_{\rm DM}$ (right panel) with the dark matter self-annihilation cross-section, for different locations of the emission region. }
	\label{fig:UpperLimitsMass}
\end{figure}

\begin{figure}[t!]
	\centering	
	\includegraphics[width=0.475\textwidth]{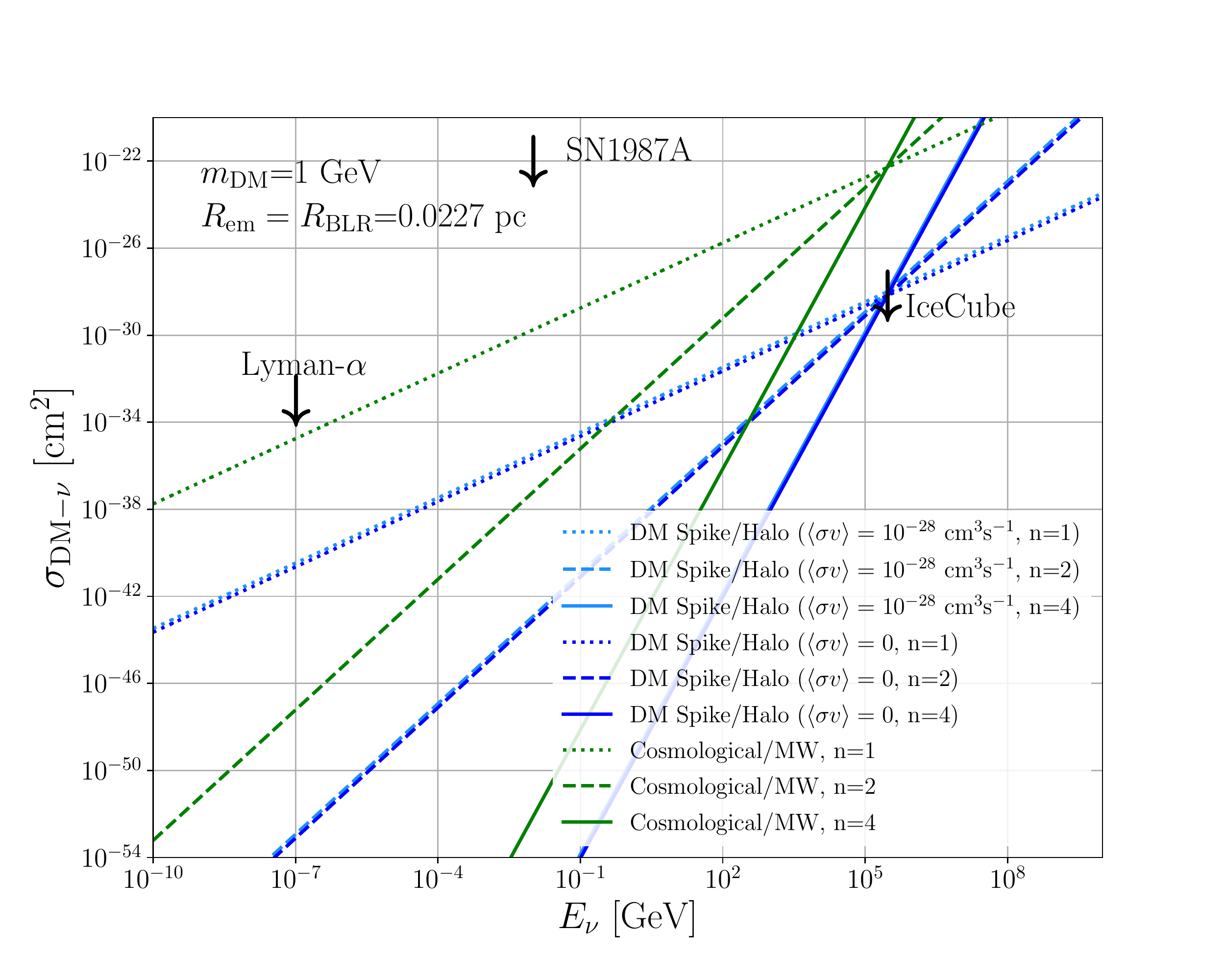}
	\includegraphics[width=0.475\textwidth]{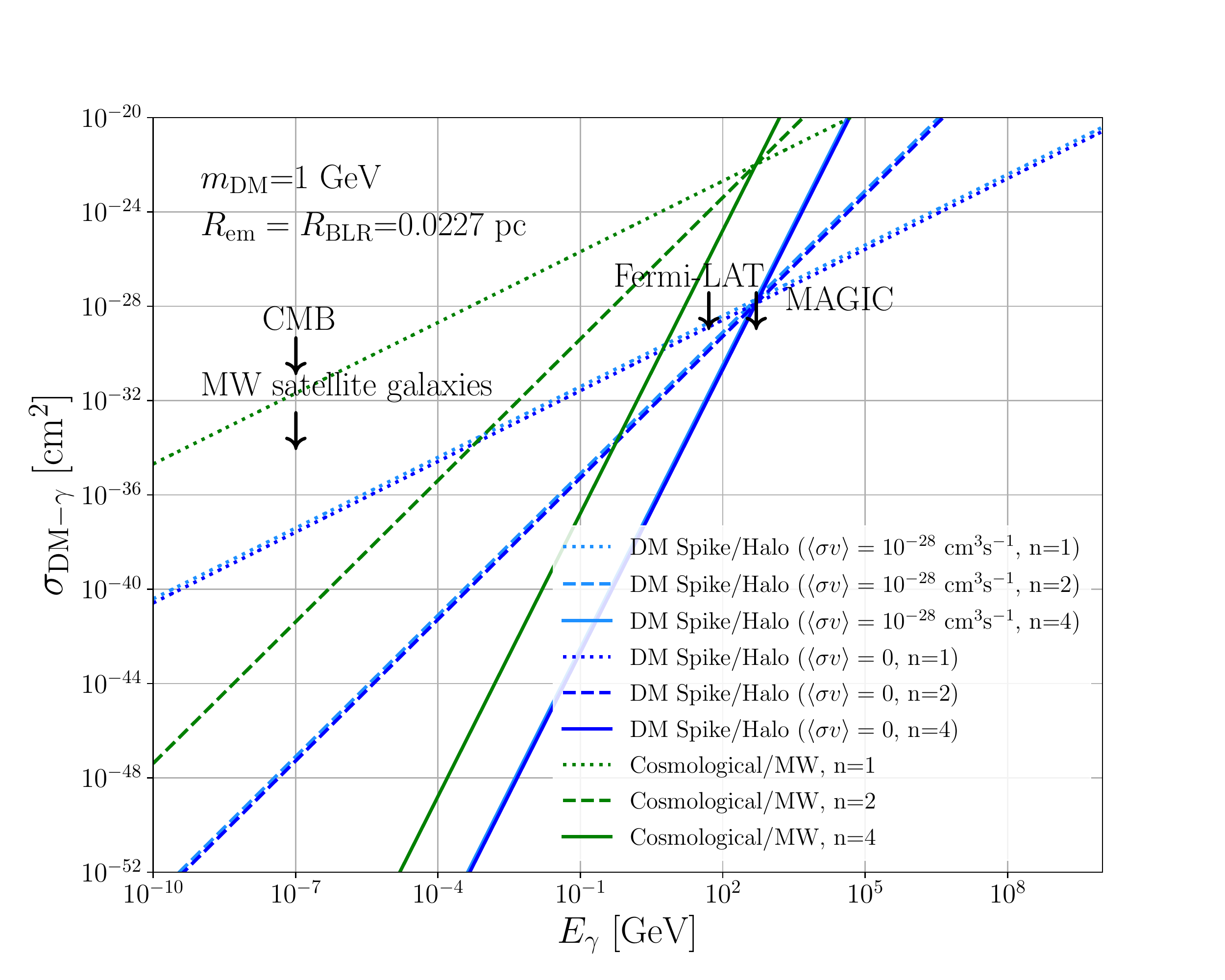}
	\centering
	\caption{Same as the top panels of Fig.\ref{fig:UpperLimitsMass}, but as a function of the energy for the case $m_{\rm DM}=1$ GeV and $R_{\rm em}=0.023$ pc, assuming that the cross-section scales with the energy as $\sigma_{\rm DM-i}=\sigma_0 (E_i/{\rm 1\, GeV})^{n}$, for $n=1,2,4$.}
	\label{fig:UpperLimitsEnergy}
\end{figure}

Our main results are shown in Figures \ref{fig:UpperLimitsMass} and \ref{fig:UpperLimitsEnergy}. The top panels of Figure \ref{fig:UpperLimitsMass} show upper limits on the dark matter-neutrino (left) and dark matter-photon (right) cross-section as a function of the dark matter mass, assuming $\gamma=1$, when the cross-section is energy independent. The blue lines are the limits derived in this work from imposing Eq.~(\ref{eq:criteria}) in different scenarios: the solid lines assume $\langle \sigma v\rangle=0$ while the dashed lines are for $\langle \sigma v\rangle=10^{-28}\,{\rm cm}^2{\rm s}^{-1}$, which amount to two different dark matter spike profiles. The different shades of blue correspond to different locations of the emitting region of neutrinos and photons. When $\langle \sigma v\rangle=0$, we find the upper limits  $\frac{\sigma_{\rm DM-\nu}}{m_{\rm DM}} \leq 2.0 \times 10^{-29}$ cm$^{2}$/GeV and $\frac{\sigma_{\rm DM-\gamma}}{m_{\rm DM}} \leq 4.1 \times 10^{-29} $ cm$^{2}$/GeV. The limits for other halo profiles can be calculated from Eqs.~(\ref{eq:Sigma}-\ref{eq:f}). Assuming $R_{\text{em}}=R_{\text{BLR}}$,
our upper limits change by a factor of at most $\sim 2$ for halo profiles in the range $\gamma=0.7-1.4$. 

As the dark matter self-annihilation cross-section increases, the effect of the flux attenuation becomes smaller, and the limits on the dark matter neutrino and photon scattering cross-sections become weaker. This is illustrated in the bottom panels in Figure~\ref{fig:UpperLimitsMass}, which show the dependence on the upper limits on $\sigma_{\rm DM-i}/m_{\rm DM}$ as a function of $\langle \sigma v\rangle/m_{\rm DM}$ for different values of the location of the emission region $R_{\rm em}$. The lines reflect the dependence of the column density on the cross-section discussed in Section \ref{sec:DMspike}: for small dark matter self-annihilation cross-sections, $\Sigma_{\rm DM}$  is roughly constant, since the halo profiles are practically indistinguishable, and thereby the limit on the scattering cross-section; for larger annihilation cross-sections,  $\Sigma_{\rm DM}$ is inversely proportional to $\langle \sigma v\rangle/m_{\rm DM}$, and therefore the limit on the scattering cross-section increases linearly with $\langle \sigma v\rangle/m_{\rm DM}$; for very large annihilation cross-sections, $\Sigma_{\rm DM}$ is dominated by the passage through the dark matter halo of the host galaxy, and again the limit on the scattering cross-section becomes independent of $\langle \sigma v\rangle /m_{\rm DM}$.

For comparison, we also show in the upper panels of the Figure \ref{fig:UpperLimitsMass} (as solid green lines) the upper limits on the dark matter-neutrino and dark matter-photon scattering cross-sections from the attenuation of neutrinos or photons due to interactions with dark matter particles in the intergalactic medium and in the Milky Way halo. As is apparent from the plot, the attenuation of the flux during the passage through the dark matter spike and the galaxy hosting the blazar is very significant. Specifically, when the neutrinos or photons are emitted at $R_{\rm em}=R_{\rm BLR}$ and the dark matter does not self-annihilate, the limits on the cross-section become about six orders of magnitude stronger than those obtained when neglecting the spike and considering just the propagation through the intergalactic medium and the Milky Way halo; when the annihilation cross-section is $\langle \sigma v\rangle =10^{-28} \,{\rm cm}^2 \,{\rm s}^{-1}$, the limits are four to six orders of magnitude stronger, depending on the dark matter mass.	

We also show (as purple lines) the constraints on the cross-section obtained from the suppression of primordial density fluctuations in the early universe, which would affect the cosmic microwave background power spectrum and the Lyman-$\alpha$ forest \cite{Mangano_2006,Wilkinson:2013kia, Wilkinson:2014ksa, Escudero:2015yka}, for the case when the cross section is constant (dotted line), and when the cross section scales as the square of the temperature of the Universe (solid line). For photons, we additionally show (as a solid orange line) the limit  derived in \cite{Escudero_2018} for $m_{\rm DM} \gtrsim$ 1 GeV from Milky Way satellite galaxy counts~\cite{B_hm_2014}~\footnote{Stronger constraints can be derived for concrete models. For instance, when the dark matter couples with similar strength to electrons and to neutrinos, the limits on the dark matter-electron cross-section from \cite{Jho:2021rmn,Zhang:2020nis,Ghosh:2021vkt,Farzan:2014gza,Das:2021lcr, Lin:2022dbl} can be translated into limits on the dark matter-neutrino cross-section. Also, when the interaction dark matter-photon is due to a dark matter millicharge, the stringent limits derived in \cite{Dvorkin_2014} would apply.}.

The cosmological limits on the dark matter-neutrino cross section are a few orders of magnitude stronger than the ones derived in this work from TXS 0506+056. However, in any realistic model, the dark matter-neutrino cross section will have a power-law dependence with the neutrino energy, $\sigma_{\rm DM-\nu}=\sigma_0(\frac{E_\nu}{\rm 1 GeV})^{n}$. Since the neutrinos observed from TXS 0506+056 are significantly more energetic ($E_{\nu} \sim 290$ TeV) than the neutrinos relevant for the Lyman-$\alpha$ bounds ($E_{\nu} \sim 100$ eV), the cross-sections involved in these two observations could be largely different. In order to compare the impact of the dark matter-neutrino (or photon) interactions, it is necessary to extrapolate the limits obtained from the TXS 0506+056 to the energy scales relevant for the Lyman-$\alpha$ forest. The same rationale holds for DM-$\gamma$ interactions.
	
This is done in Figure \ref{fig:UpperLimitsEnergy}, which shows the upper limit on the dark matter neutrino (left) and dark matter-photon (right) cross-section as a function on the energy, when the cross-section scales with the energy as $\sigma_{\rm DM-\nu}=\sigma_0(E_\nu/{\rm 1\, GeV})^{n}$, for $n=1,2,4$ (dotted, dashed and solid lines, respectively). For the plot we took for concreteness $m_{\rm DM}=1$ GeV, and two spike profiles corresponding to a scenario of asymmetric dark matter ($\langle \sigma v\rangle$=0, dark blue) and of self-annihilating dark matter ($\langle\sigma v\rangle=10^{-28} \,{\rm cm}^2 \,{\rm s}^{-1}$), light blue. For comparison, we also show complementary constraints at different neutrino energies from SN1987A \cite{1996slfp.book.....R}, and from Lyman-$\alpha$ observations. As can be seen from the figure, our limits from the attenuation of the flux at the spike of TXS 0506+056 are substantially stronger than previous limits. For instance, for $n=1$ our limits on the dark matter-neutrino cross-section are $\sim$ 8 orders of magnitude stronger than those from Lyman-$\alpha$ observations, and can be even $\sim$ 20 orders of magnitude stronger for $n=2$. Similar conclusions hold for the dark matter-photon cross-section.

\section{Conclusions}\label{sec:conclusions}

High-energy neutrinos and photons have been detected from the blazar TXS 0506+056, with fluxes which are well compatible with astrophysical models. This indicates that the fluxes are not significantly attenuated by putative interactions of neutrinos or photons with the dark matter particles existing between their production point and the Earth. In this work we have investigated the possible attenuation of the fluxes due to the dark matter spike around the black hole of TXS 0506+056.

For scenarios where dark matter particles do not self-annihilate, we find the upper limits $\frac{\sigma_{\rm DM-\nu}}{m_{\rm DM}} \leq 2.0 \times 10^{-29}$ cm$^{2}$/GeV and $\frac{\sigma_{\rm DM-\gamma}}{m_{\rm DM}} \leq 4.1 \times 10^{-29} $ cm$^{2}$/GeV, which are $\sim$ 7 orders of magnitude stronger than constraints derived in previous works from the attenuation of the fluxes in the intergalactic medium and the Milky Way halo. Assuming that the cross-section is independent of the energy, the limits on the dark-matter neutrino (dark matter-photon) cross-section are $\sim$ 5 ($\sim$ 2) orders of magnitude weaker than those stemming from the cosmic microwave background and from large scale structure. Similar conclusions hold when the dark matter particles can self-annihilate, although in this case the constraints become weaker, due to the flattening of the spike at small distances from the black hole. 

We have also considered scenarios where the scattering cross-section depends with the energy as a power-law $\sigma_{\rm DM-i}=\sigma_0(\frac{E_\nu}{\rm 1 GeV})^{n}$, with $n=1,2$ or $4$. In this case the constraints on the dark matter-neutrino and dark matter-photon cross section from the attenuation in the spike of TXS 0506+056 are several orders of magnitude more stringent than those from Cosmology. 

Blazar observations in neutrinos and gamma-rays therefore constitute a powerful probe of dark matter interactions, especially for scenarios where the cross-section is energy dependent. The likely discovery of more and more neutrino sources in current and future neutrino telescopes, and their identification with gamma-ray sources, will provide very valuable information about the dark matter microphysics, and perhaps provide hints for neutrino or photon interactions with dark matter particles.

\subsection*{Acknowledgments}

We are grateful to Sergio Palomares-Ruiz, Miguel Escudero, Elisa Resconi, Manel Errando and James H. Buckley for useful discussions. This work was supported by the Collaborative Research Center SFB1258 and by the Deutsche Forschungsgemeinschaft (DFG, German Research Foundation) under Germany's Excellence Strategy - EXC-2094 - 390783311. The work of FF was supported in part by the U.S. Department of Energy under Grant No. DE-SC0017987.

\subsection*{Note added}

While this paper was being finalized, Ref. \cite{Cline:2022qld} appeared also discussing the attenuation of the neutrino flux in the dark matter spike of blazars.

\printbibliography

\end{document}